\pdfoutput=1
\documentclass{article}

\PassOptionsToPackage{numbers, compress}{natbib}

 \usepackage[preprint]{neurips_2022}

\usepackage[utf8]{inputenc} 
\usepackage[T1]{fontenc}    
\usepackage{hyperref}       
\usepackage{url}            
\usepackage{booktabs}       
\usepackage{amsfonts}       
\usepackage{amsmath}
\usepackage{nicefrac}       
\usepackage{microtype}      
\usepackage{xcolor}         
\usepackage{bm}
\usepackage{subcaption}
\usepackage{braket}
\usepackage{float}
\usepackage{graphicx}
\usepackage{wrapfig}
\title{Physics-Informed Neural Networks as Solvers for the Time-Dependent Schr\"odinger Equation}

\author{%
   Karan Shah \\
  Center for Advanced Systems Understanding\\
  Helmholtz-Zentrum Dresden-Rossendorf\\
  G\"orlitz, Germany \\
  \texttt{k.shah@hzdr.de} \\
  \And
   Patrick Stiller \\
  Institute of Radiation Physics\\
  Helmholtz-Zentrum Dresden-Rossendorf\\
  Dresden, Germany \\
  \texttt{p.stiller@hzdr.de} \\
  \And
   Nico Hoffmann \\
  Institute of Radiation Physics\\
  Helmholtz-Zentrum Dresden-Rossendorf\\
  Dresden, Germany \\
  \texttt{n.hoffmann@hzdr.de} \\
   \And
   Attila Cangi \\
  Center for Advanced Systems Understanding\\
  Helmholtz-Zentrum Dresden-Rossendorf\\
  G\"orlitz, Germany \\
  \texttt{a.cangi@hzdr.de} \\
}

\begin{document}

\maketitle

\begin{abstract}
  We demonstrate the utility of physics-informed neural networks (PINNs) as solvers for the non-relativistic, time-dependent Schrödinger equation. We study the performance and generalisability of PINN solvers on the time evolution of a quantum harmonic oscillator across varying system parameters, domains, and energy states.
\end{abstract}

\section{Introduction}
\label{sec:introduction}
In recent years, there was a surge in the use of machine learning (ML) techniques in the physical sciences \cite{RevModPhys.91.045002}, referred to as scientific machine learning. This has led to the rise of physics-informed machine learning \cite{karniadakisPhysicsinformedMachineLearning2021a}, where physics-based constraints are used to guide ML models, achieved by incorporating structured prior information derived from physical laws into the learning algorithm.

Given the importance of numerical simulations across scientific disciplines, many ML surrogates for solving differential equations on a large scale \cite{https://doi.org/10.1002/gamm.202100006} have been developed. These approaches include the deep Galerkin method \cite{sirignanoDGMDeepLearning2018}, PINNs \cite{raissiPhysicsinformedNeuralNetworks2019c}, neural operators \cite{li2021fourier}, and DeepONets \cite{luLearningNonlinearOperators2021}.

PINNs are a class of ML algorithms for solving forward and inverse problems that are represented by partial differential equations. As opposed to numerical solvers where a new solution must be computed whenever there is a change in the domain or system parameters, PINNs offer a mesh-free alternative where the solver can be used for systems on arbitrary grid resolutions and system parameters at a fixed inference cost \cite{markidisOldNewCan2021a}. While training the PINN models can be a substantial computational cost initially, it can be amortized over time due to rapid inference over a wider set of system parameters. However, PINNs also have demonstrated shortcomings such as spectral bias against high-frequency solutions \cite{wangWhenWhyPINNs2020}, overfitting to trivial solutions \cite{leiteritzHowAvoidTrivial2021}, and poor performance for systems with shocks \cite{Fuks_2020} and for larger time domains \cite{10.1007/978-3-031-08754-7_45}. The theory and applications of PINNs are active areas of research, with developments such as the RNN-DCT-PINN \cite{10.1007/978-3-031-08754-7_45}, the gated-PINN \cite{stillerLargeScaleNeuralSolvers2020}, and the variational-PINN \cite{Kharazmi2019VariationalPN} which increase their performance and scalability. Popular ML frameworks for PINN models include Neural Solvers \cite{stillerLargeScaleNeuralSolvers2020}, Modulus \cite{hennighNVIDIASimNetTM2020}, and DeepXDE \cite{lu2021deepxde}.

The non-relativistic Schr\"odinger equation is the fundamental equation for describing quantum systems. The wavefunction $\phi$, from which all observables of a system can be calculated, is obtained by solving it. We distinguish two classes of problems: the time-independent Schr\"odinger equation (TISE) yields the wave function $\phi$ and the associated energies for a static system in terms of an eigenvalue problem, while  
the time-dependent Schr\"odinger equation (TDSE) describes the dynamics of a quantum system, i.e., the time evolution of $\phi$. Prior works have dealt with using ML to solve the SE. These include models such as fully-connected networks (FCN) \cite{lagarisArtificialNeuralNetwork1997b, raduNeuralNetworkApproaches2022}, reservoir computing \cite{domingoAdaptingReservoirComputing2022}, PINNs \cite{664609} for the eigenvalue problem defined by the TISE, and residual networks \cite{xieMachineLearningMethodTimeDependent2021} and LSTMs \cite{yangConditionalSeq2SeqModel2022} for the TDSE. The TDSE has also been used as benchmark for PINNs \cite{raissiPhysicsinformedNeuralNetworks2019c, karniadakisPhysicsinformedMachineLearning2021a, stillerLargeScaleNeuralSolvers2020}. However, in this work, we go beyond prior investigations by examining the utility of PINN solvers for the TDSE across varying system parameters, domains, and energy states. 

\section{Methods}
\label{sec:methods}
\subsection{PINNs}
PINNs are constructed by encoding the constraints posed by a given differential equation and its boundary conditions into the loss function of a deep learning model, usually, an FCN. These constraints guide the network to finding a solution to the differential equation.

A partial differential equation (PDE) is defined by $f$ with solution $u(\mathbf{x},t)$ governed by 
\begin{align}\label{eq:pde}
f(u) &:=u_{t}+\mathcal{N}[u;\lambda],~ \bm{x} \in \Omega,~ t \in [0,T] \,, \qquad f(u) = 0 \ ,
\end{align} 
where $\mathcal{N}[u;\lambda]$ is a differential operator parameterised by $\lambda$, $\Omega \in \mathbb{R^D}$, and $\bm{x} = (x_1,x_2,...,x_d)$ \\ with boundary conditions $\mathcal{B}(u, \bm{x},t)=0  \text { on } \partial \Omega \ $
 and initial conditions $\mathcal{T}(u, \bm{x},t)=0  \text { at }  t = 0 \ $.

A neural network $u_{net}: \mathbb{R}^{D+1}\mapsto \mathbb{R}^{1}$ is constructed as a surrogate model $f_{net}=f(u_{net})$ for the true solution $u$.
Constraints are  encoded in the loss term $L$ for neural network optimization
\begin{equation}
L={\lambda_f L_{f}}+{\lambda_{BC} L_{BC}}+{\lambda_{IC} L_{IC}}\ ,
\label{eq:pinn_loss}
\end{equation} with $\lambda_f, \lambda_{BC}, \lambda_{IC}$ being the regularization parameters.
The PDE loss $L_{f} = (1/N_{f}) \sum_{i=1}^{N_{f}}\left|f_{net}\left(\bm{x}_{f}^{i}, t_{f}^{i}\right)\right|^{2}$ denotes the error in the solution within the interior points of the system and is calculated for $N_f$ collocation points. The boundary loss $L_{BC} =(1/N_{BC}) \sum_{i=1}^{N_{BC}}\left|u_{net}\left(\bm{x}_{BC}^{i}, t_{BC}^{i}\right)-u\left(\bm{x}_{BC}^{i}, t_{BC}^{i}\right)\right|^{2}$ is the constraint imposed by the boundary conditions, whereas $L_{IC} =(1/N_{IC}) \sum_{i=1}^{N_{IC}}\left|u_{net}\left(\bm{x}_{IC}^{i}, t_{IC}^{i}\right)-u\left(\bm{x}_{IC}^{i}, t_{IC}^{i}\right)\right|^{2}$
imposes the initial conditions. Both of them are calculated on a set of $N_{BC}$ boundary points and $N_{IC}$ initial points, respectively, where $u_{net}$ refers to the approximate solution predicted by the PINN and $u$  denotes the true value according to the boundary and initial conditions at that point. The distribution of collocation points is visualized in Fig. \ref{fig:pinn_colloc}.

Once trained, the neural network is used to solve the PDE, potentially for a range of parameters $\lambda$ \cite{markidisOldNewCan2021a}.

\subsection{Time Dependent Schr\"odinger Equation}
A PINN is constructed for solving the TDSE
\begin{equation}
i \frac{\partial \phi(\mathbf{r}, t)}{\partial t}-\hat{H} \phi(\mathbf{r}, t)=0\,,
\end{equation} 
where $\hat{H}$ denotes the Hamiltonian of the problem and $\phi(\mathbf{r}, t)$ the solution which is commonly referred to as a wave function that depends on a spatial coordinate $\mathbf{r}$ in three-dimensional space and time $t$. Note that we adopt Hartree atomic units, i.e., $\hbar=k_\textnormal{B}=m_e=1$, so energies are measured in Hartree and lengths in Bohr radii. 
The Hamiltonian represents the specific problem, such as the kinetic and potential energies of particle species and their interactions among each other and with external fields.

In this conceptual work, we resort to a simple but archetypical model problem, namely non-interacting and spinless particles in a quantum harmonic oscillator in one spatial dimension with Hamiltonian
\begin{equation}
\hat{H}_x=-\frac{1}{2}\frac{\partial^{2}}{\partial x^{2}}+\frac{\omega^{2}}{2}x^{2}\,,
\end{equation}
where $\omega$ denotes the frequency of the harmonic oscillator.
The most fundamental kind of quantum dynamics is achieved by a superposition of eigenstates which are solutions of the corresponding TISE. The analytical eigenstates of the harmonic oscillator are $\phi_n(x) = \phi_0(x)(2^n n!)^{-1/2}H_n(\sqrt{\omega}x)$ and the corresponding eigenvalues are $\epsilon_n = \omega(n+1/2)$ with $n \in \mathbb{N}$, where $H_n$ denotes the Hermite polynomials and $\phi_0(x) = (\omega/\pi)^{1/4}\exp\left(-\omega x^2/2\right)$.
Superimposing two eigenstates $\phi_{m,n}(x,t) = \left[ e^{-i \epsilon_m t}\phi_m(x) + e^{-i \epsilon_n t}\phi_n(x)\right]/\sqrt{2}$ yields a solution of the TDSE, where the entire dynamics takes place due to the phases $\exp\left(-i \epsilon_n t\right)$.
Since the neural network is constrained to $\mathbb{R}$, the complex-valued solutions can be represented as $\phi(x,t) = u + iv$, where $u = \operatorname{Re}(\phi)$ is the real part and $v=\operatorname{Im}(\phi)$ the imaginary part.
The TDSE can then be written as in terms of $u$ and $v$ as
\begin{equation}
    \left(-\frac{\partial v}{\partial t}+\frac{1}{2}\frac{\partial^{2} u}{\partial x^{2}}-\frac{\omega^{2}}{2}x^{2}\right) + i \left(\frac{\partial u}{\partial t}+\frac{1}{2}\frac{\partial^{2} v}{\partial x^{2}}-\frac{\omega^{2}}{2}x^{2}\right) = 0 \ .
\end{equation}
Consequently, a PINN was constructed with inputs $(x,t,\omega)$ and outputs $(u_{net},v_{net})$. The error of the predictions was quantified on the probability density $|\phi(x,t)|^2 = u(x,t)^2 + v(x,t)^2$.


\section{Results}
\label{sec:results}
The domains for the TDSE of the quantum harmonic oscillator are $x \in [-\pi, \pi]$ and $t \in [0, 2\pi]$ with Dirichlet boundary conditions $u(x_0,t)=0,~ v(x_0,t)=0$ for $x_0 \in \{-\pi,\pi\}$ and initial conditions $u(x,0) = \operatorname{Re}\left[\phi_{m,n}(x,0)\right] = \phi_{m,n}(x,0),~ v(x,0)=\operatorname{Im}\left[\phi_{m,n}(x,0)\right]=0 $.

\begin{table}[H]
  \caption{Results. (Here FP denotes FCN-PINN.)}
  \label{result-table}
  \centering
  \begin{tabular}{lllll}
    \toprule

    System & Solver     & MSE $|\phi|^2$     & Training & Inference \\
     &      &    &  time (min) &  time (s) \\
    \midrule
    Baseline   & FP  & 2.94e-5   &   16.12  &  4.26\\

    \midrule
    Generalisability   & FP (Interpolation)   & (2.72$\pm$2.18)e-5  &   130.72  & (9.79$\pm$0.10) \\
       & FP (Extrapolation)   &  (2.65$\pm$1.56)e-3 &   &   \\
    \midrule
    Larger time domain  & FP &  4.38e-2 &  57.56 &  12.25  \\
        & FP (Causal) &  3.62e-3 &    63.33  &  12.35 \\
    \midrule
    Higher energy states  &  FP &  1.25e-2 & 29.73 &  3.98      \\
        & FP (Causal) & 1.49e-3   & 85.85 &  4.28    \\
    \bottomrule
  \end{tabular}
\end{table}

\subsection{Baseline}
\label{sec:baseline}
The initial surrogate model was created for a one-dimensional quantum harmonic oscillator with a fixed parameter $\omega = 1$. The initial state was chosen as a superposition of the ground and the first excited states $\phi_{0,1}(x,t) = \left[ e^{-i \epsilon_0 t}\phi_0(x) + e^{-i \epsilon_1 t}\phi_1(x)\right]/\sqrt{2}$.
\begin{figure}[H]
\centering
\includegraphics[width=0.75\textwidth]{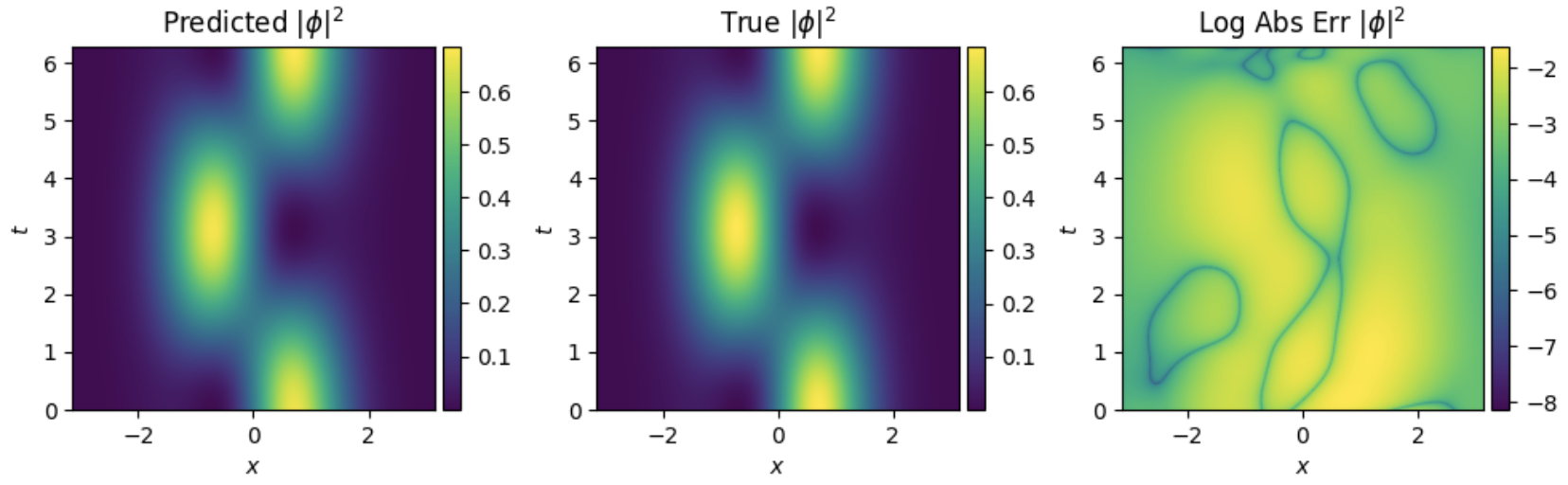}
\caption{True and predicted values for density $|\phi|^2$ for $\phi_{0,1}(x,t)$ with $\omega$ = 1.0.}
\label{fig:1d_qho_0_1_snapshot}
\end{figure}

Fig. \ref{fig:1d_qho_0_1_snapshot} shows the results for the baseline model. It illustrates how the probability density of the superpositions state propagates in time, where the left plot shows the prediction of the PINN and the center plot the ground truth.  As shown, on the right plot, the PINN solver yields the structure of the solution very accurately. The spatially and temporally resolved MSEs are in the order of $10^{-5}$.


\subsection{Generalisability}
\label{sec:generalisability}
\begin{figure}
\centering
\begin{minipage}{.31\textwidth}
  \centering
  \includegraphics[width=\linewidth]{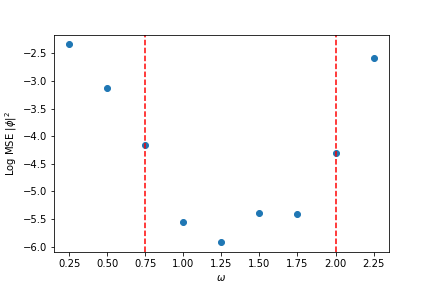}
  \captionof{figure}{Log MSE $|\phi|^2$ for PINN solutions over varying values of $\omega$. Training range within red dashed lines.}
  \label{fig:1d_qho_0_1_omega_variance}
\end{minipage}%
\hfill
\begin{minipage}{.31\textwidth}
  \centering
  \includegraphics[width=\linewidth]{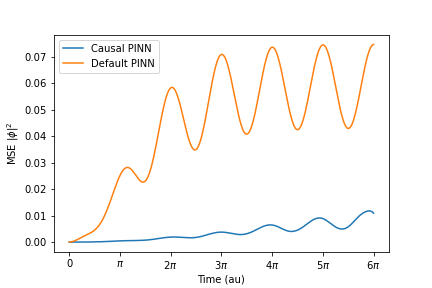}
  \captionof{figure}{MSE $|\phi|^2$ over time for large time domain.}
  \label{fig:1d_qho_0_1_6_pi_error}
\end{minipage}
\hfill
\begin{minipage}{.31\textwidth}
  \centering
  \includegraphics[width=\linewidth]{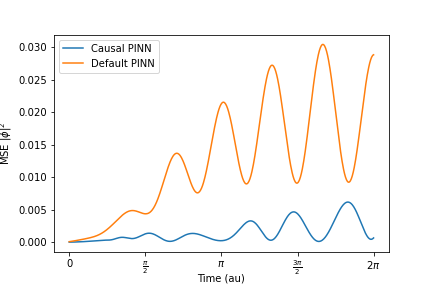}
  \captionof{figure}{MSE $|\phi|^2$ over time for higher energy states.}
  \label{fig:1d_qho_0_3_error}
\end{minipage}
\end{figure}

To demonstrate the performance over changing parameters, the PINN was trained for the same initial state as the baseline in a range of values $\omega \in [0.75,2.0]$. After training, the inference was carried out for $\omega \in [0.5,2.5]$. As illustrated in Fig.~\ref{fig:1d_qho_0_1_omega_variance}, good performance was observed on previously unseen $\omega$ values within the range of the training range (interpolation) but accuracy diminished somewhat as the values of $\omega$ deviated from the training range during inference (extrapolation). This can be attributed to the change in the width of the wave function relative to our fixed space domain. The width is inversely proportional to $\omega$. For lower values of $\omega$, the width is too broad and violates the fixed domain boundary conditions. For higher values of $\omega$, the width is too narrow and the network underfits to a zero-valued solution. This problem could be alleviated by using dynamic domains while training PINNs, such that the domain adjusts to the width of the wave function. 

\subsection{Larger time domains}
\label{sec:lt_domain}
Next, a longer time domain $t \in [0, 6\pi]$ for the baseline $\phi_{0,1}(x,t)$ system was studied. The FCN-PINN resulted in a high mean error in the order of $10^{-2}$ with a loss of structure at later time steps as shown in Fig.~ \ref{fig:1d_qho_0_1_6_pi_error} (orange curve). To mitigate this, the loss function was modified to take causality into account \cite{Wang2022RespectingCI}. This was done by splitting the time domain into equally spaced components, with the loss for each component being a weighted cumulative sum of $L_f$ of the domains up to that component. This enforces causality in the time domain. With this modification, the error is reduced to $10^{-3}$ with the correct structure preserved throughout the time domain (blue curve).

\subsection{Higher energy states}
\label{sec:he_states}
As a demonstration for higher energy states, the FCN-PINN was trained on the initial state $\phi_{0,3}(x,t) = \left[ e^{-i \epsilon_0 t}\phi_0(x) + e^{-i \epsilon_3 t}\phi_3(x)\right]/\sqrt{2}$ .
The predicted solution with the default FCN architecture falsely converged to the static ground state $\phi_0$ as time evolved. As can be seen from Fig.~\ref{fig:1d_qho_0_3_error}, the error increases significantly for later time steps (orange curve). Subsequently, a PINN with a loss function that preserves causality, as explained in Section \ref{sec:lt_domain}, was used. This solves the problem of false convergence. It captures the structure of the solution over a long period of time evolution with a low MSE on the order of $10^{-3}$ (blue curve).

\section{Future Work}
There has been a rise in the applications of machine learning in computational chemistry and materials science \cite{PhysRevMaterials.6.040301}. Density functional theory (DFT) is the most widely used method for computing chemical and material properties. Hence, DFT calculations utilize a considerable proportion of computational resources on academic HPC systems. The underlying equations for time-dependent DFT are the Schr\"odinger-like time-dependent Kohn-Sham (TDKS) equations:
\begin{equation}
\left[-\frac{1}{2}\nabla^2 + v_s[n](\bm{r},t)\right]\phi_j(\bm{r},t) = i \frac{\partial \phi_j(\bm{r},t)}{\partial t} 
\label{eq:KSequation.TD}
\end{equation}
where the electronic density 
$ n(\bm{r}, t) = \sum_j \, |\phi_j(\bm{r}, t)|^2 $ is the quantity of interest. The Kohn-Sham potential $v_s[n](\bm{r},t)$ is a functional of the density $n$ which is a function of position $\bm{r}$ and time $t$. These coupled equations are solved through iterative numerical algorithms which represent a significant computational cost in the TDDFT workflow\cite{doi:10.1063/1.1774980, gomezpueyoPropagatorsTimeDependentKohn2018}. Given the effectiveness of PINNs for the TDSE, the next step is to use PINNs to accelerate TDKS calculations.

\section{Conclusion}
\label{sec:conclusion}
PINNs act as surrogate models for solving the TDSE. The primary advantage of PINNs over traditional numerical solvers is that they are mesh-free and can be used to perform simulations at arbitrary resolutions once sufficiently trained. Depending upon the training regiment, PINNs can be generalized to a range of the PDE parameter space. Their mesh-free nature and generalisability can be utilized for accelerating simulations of electron dynamics. While FCN-PINNs are not sufficient to reproduce complex dynamics, this can be alleviated with techniques such as causal training. Extending this work to TDDFT, a PINN framework would enable on-the-fly modeling of the electronic response properties of laser-excited or shock-compressed samples in various scattering experiments that are conducted at photon sources around the globe. This would enable fast simulations that generalize well over the input parameters of the experimental setup.

\section{Broader Impact}
While this work is at a preliminary stage, we believe that accelerating TDDFT calculations is a net benefit. Accelerating TDDFT calculations has a broad impact on the fields such as computational chemistry and materials science with applications in myriad domains such as materials design, drug discovery, and green energy generation. Results obtained in this line of research can also be used to accelerate numerical simulations in other fields of science and engineering.

\begin{ack}
This work was supported by the Center for Advanced Systems Understanding (CASUS) which is financed by Germany’s Federal Ministry of Education and Research (BMBF) and by the Saxon state government out of the State budget approved by the Saxon State Parliament. 
\end{ack}

\medskip
\bibliography{main}

\appendix

\section{Computation Details}
\subsection{Network architecture and hyperparameters}
\label{sec:arch}

\subsubsection{Sampling} 
We used a batch size of 3140 points for the interior, 200 points for the boundary conditions and 314 points for initial conditions. 

\subsubsection{Architecture}

\begin{figure}[H]

\centering
\begin{subfigure}{0.7\textwidth}
    \includegraphics[width=1.0\textwidth]{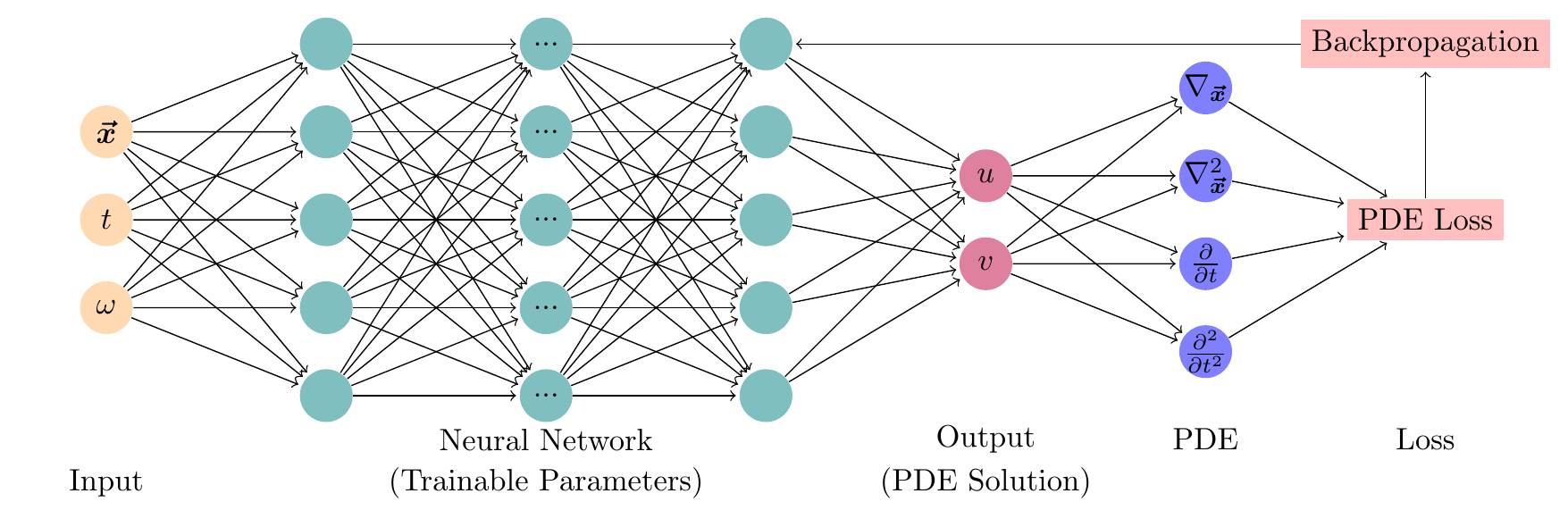}
    \caption{Common architecture of PINNs.}
    \label{fig:pinn_architecture}
\end{subfigure}
\hfill
\begin{subfigure}{0.25\textwidth}
    \includegraphics[width=1.0\textwidth]{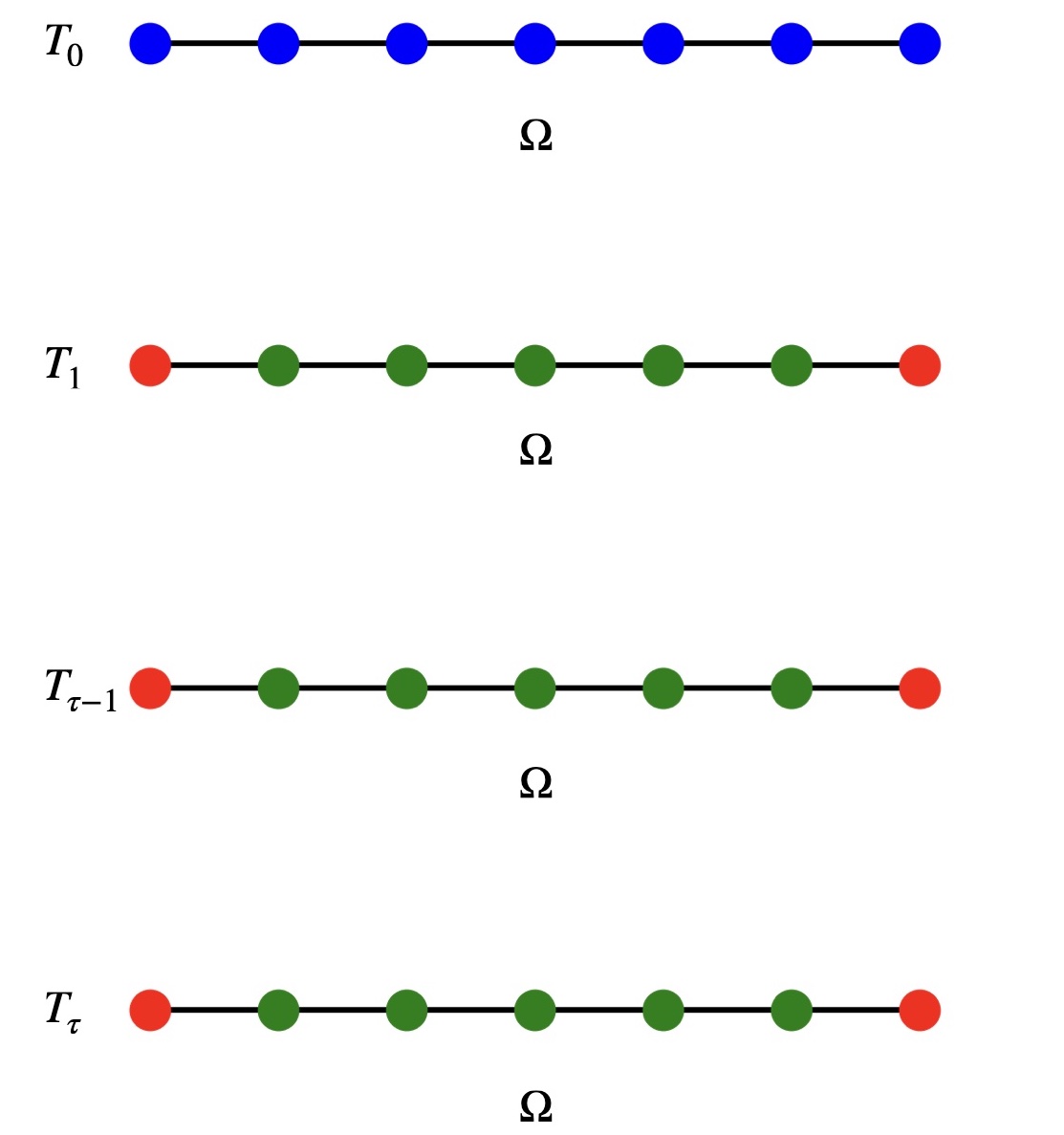}
    \caption{Distribution of collocation points for calculating the loss function in Eq (\ref{eq:pinn_loss}).}
    \label{fig:pinn_colloc}
\end{subfigure}

\caption{PINN Architecture}
\label{fig:pinn_details}

\end{figure}

Because of the derivatives needed in the PINN loss, the activation function has to be $k+1$ differentiable for a PDE of order $k$. This restricts the choice of activation functions to functions like tanh and SiLU.

FCN-PINN (Sections \ref{sec:baseline}, \ref{sec:generalisability}, \ref{sec:lt_domain}, \ref{sec:he_states}): We used a fully-connected network with 6 layers consisting of 512 neurons each. The activation function used was tanh. We calculate the loss function as described in Eq. \ref{eq:pinn_loss}. We use the ADAM optimizer with $\beta_1 =0.09, \beta_2=0.999$ . The learning rate is initialized at $\alpha_0 = 0.001$ with exponential decay rate $\gamma = 0.9$ at decay steps $t_{\gamma} = 2000$ training steps with schedule $\alpha_t = \alpha_0 \gamma^{\frac{t}{t_{\gamma}}}$.

For the generalisability study (Section \ref{sec:generalisability}), we used a fully-connected network with 12 layers consisting of 512 neurons each.

Causal-PINN (Sections \ref{sec:lt_domain}, \ref{sec:he_states}): We used a fully-connected network with 6 layers consisting of 512 neurons each. The activation function used was SiLU. We calculate the loss function with the causal training scheme described in \cite{Wang2022RespectingCI}. We use the ADAM optimizer with $\beta_1 =0.09, \beta_2=0.999$ . The learning rate is initialized at $\alpha_0 = 0.001$ with exponential decay rate $\gamma = 0.95$ at decay steps $t_{\gamma} = 100$ training steps with schedule $\alpha_t = \alpha_0 \gamma^{\frac{t}{t_{\gamma}}}$.

In both cases, training is continued till convergence is reached for MSE $|\phi|^2$ or maximum number of training iterations for each experiment.

\subsubsection{Evaluation}
The PINN solution $u_{net}(x,t), v_{net}(x,t)$ is used to calculate the density $|\phi_{net}(x,t)|^2 = u_{net}^2(x,t) + v_{net}^2(x,t)$ on a grid with resolution $\Delta x, \Delta t =0.01 \text{au}$ in the domain specified for that system, i.e. $628 \times 628$ uniformly-spaced points in Sections \ref{sec:baseline}, \ref{sec:generalisability}, \ref{sec:he_states} and $1884 \times 628$ points in Section \ref{sec:lt_domain}.
\begin{equation}
MSE |\phi_{net}(x,t)|^2 = \frac{1}{N}\sum_{i=1}^{N}\left||\phi(x_i,t_i)|^2 - |\phi_{net}(x_i,t_i)|^2 \right|^2 \ ,
\end{equation} calculated over all $N$ points of the evaluation grid, where $\phi(x_i,t_i)$ is the analytical solution.

\subsection{Software and Hardware}
\label{sec:HWSW}
\subsubsection{Software}
We utilise the Modulus PINN framework \cite{hennighNVIDIASimNetTM2020} and PyTorch ML framework  \cite{NEURIPS2019_9015} in this work.

\subsubsection{Hardware}
All experiments were carried out on nodes consisting of one 12 Core Intel Xeon 3.0 GHz CPU and one Nvidia Tesla V100 (32 GB) GPU.

HPC Usage: We estimate total usage to be 18  GPU-hours for the experiments in this paper.

\section{Analytical Solutions}

The analytical solution for the baseline is
\begin{equation}
\phi_{0,1}(x,t) = \frac{1}{\sqrt{2}}\sqrt[4]{\frac{\omega}{\pi}}\exp\left(-\frac{\omega x^2}{2}\right)\left( exp{\left(-i \frac{\omega}{2} t\right)}+exp{\left(-i \frac{3\omega}{2}  t\right)}\sqrt{2\omega}x\right) .
\end{equation}

The analytical solution in higher energy state is
\begin{equation}
    \phi_{0,3}(x,t) = \frac{1}{\sqrt{2}}\sqrt[4]{\frac{\omega}{\pi}}\exp\left(-\frac{\omega x^2}{2}\right)\left( exp{\left(-i \frac{\omega}{2} t\right)}+
exp{\left(-i \frac{7\omega}{2} t\right)}\frac{1}{\sqrt{3}}\left(2\sqrt{\omega^3}x^3 - 3\sqrt{\omega}x\right)\right).
\end{equation}

\end{document}